\documentclass[twocolumn,superscriptaddress,showpacs,amsfonts,amsmath,amssymb,prb]{revtex4}
\usepackage{graphicx}

\begin{document}

\title{Contact induced spin relaxation in Hanle spin precession measurements}

\author{T. Maassen}
	\email{t.maassen@rug.nl}
\author{I. J. Vera-Marun}
\author{M. H. D. Guimar\~{a}es} 
\author{B. J. van Wees}
	\affiliation{Physics of Nanodevices, Zernike Institute for Advanced Materials, University of Groningen, Groningen, The Netherlands}

\date{\today}

\begin{abstract}
In the field of spintronics the ``conductivity mismatch'' problem remains an important issue. Here the difference between the resistance of ferromagnetic electrodes and a (high resistive) transport channel causes injected spins to be backscattered into the leads and to lose their spin information. We study the effect of the resulting contact induced spin relaxation on spin transport, in particular on non-local Hanle precession measurements. As the Hanle line shape is modified by the contact induced effects, the fits to Hanle curves can result in incorrectly determined spin transport properties of the transport channel. We quantify this effect that mimics a decrease of the spin relaxation time of the channel reaching more than 4 orders of magnitude and a minor increase of the diffusion coefficient by less than a factor of 2. Then we compare the results to spin transport measurements on graphene from the literature. We further point out guidelines for a Hanle precession fitting procedure that allows to reliably extract spin transport properties from measurements.
\end{abstract}
 \maketitle
%75.76.+j Spin transport (magnetoelectronics)
% Spin diffusion, 75.40.Gb
%  spin polarized transport in semiconductors, 72.25.Dc
% 72.80.Vp Electronic transport graphene
%%%%%%%%%%%%%%%%%%%%%%%%%%%%%%%%%%%%%%%%%%%%%%%%%%%%%%%%%%%%%%%%%%%%%%%%%%%%%%%%%%%%%%%%%%%%%%%%%%%%%%%%%%%%%%%%%%%%%%%%%%%%%%%%%%%%%%%%%%%%%%%%%%%%%%%%%%%%%
\section{\label{sec:Introduction}Introduction}
New concepts like the spin transfer torque, the transport of spin information over long distances and the prospect of spin field effect transistors keep spintronics an inspiring field \cite{IJRD50_Wolf2006, APS57_Fabian2007}. But before new types of spintronic devices can be build we need both materials that efficiently generate spin currents as injector and detector electrodes and materials with long spin relaxation lengths ($\lambda$) and times ($\tau$) to transport the spins with only little losses.\\
While ferromagnetic metals can spin polarize currents and are therefore used to inject and detect spins, semiconductors offer low spin relaxation which make them good candidates to be used as transport channels. One of the main challenges when combining the two types of materials is the ``conductivity mismatch'' problem \cite{PRB62_Schmidt2000, PRB62_Rashba2000}. As the electrical resistance in the ferromagnetic electrodes is in general lower than in the semiconducting transport channel, the injected spins tend to be reabsorbed by the leads and loose their spin orientation. \\
Graphene being an intermediate between metal and semiconductor systems is a prototype example for the conductivity mismatch, as graphene based devices can be well described following simple spin diffusion theory. Here, the long spin relaxation lengths of several $\mu m$ measured at room temperature are already promising \cite{N448_Tombros2007}, but still stay behind the theoretical prospects based on the high mobilities combined with weak spin orbit coupling and low hyperfine interactions \cite{TEPJ148_Huertas-Hernando2007}. While some research aims to understand the spin relaxation mechanism in graphene \cite{PRB80_Jozsa2009, PRB80_Ertler2009, PRL103_Castro2009, PRL107_Yang2011, NJoP14_Zhang2012, PRL108_Ochoa2012} and to understand the influence of the direct environment of the graphene transport channel \cite{PRL104_Pi2010, PRB83_Maassen2011, NL12_Maassen2012, NL12_Guimaraes2012, c_McCreary2012, c_Zomer2012} the conductivity mismatch can play an important role in the origin of spin relaxation in the measured devices. To prevent this mismatch high resistive barriers between the contacts and the graphene channel are included \cite{PRB62_Rashba2000, PRB80_Popinciuc2009, PRL105_Han2010, APL97_Dlubak2010, NPaoP_Dlubak2012, JoVSTBMaNS30_Abel2012}. \\
The most common and reliable way to probe spin transport properties is by performing measurements in the non-local spin-valve geometry \cite{N416_Jedema2002, N448_Tombros2007, APS57_Fabian2007} because it enables to separate spin and charge currents, avoiding spurious effects \cite{PRB73_Tombros2006}. 
Popinciuc et al.\cite{PRB80_Popinciuc2009} describe, in agreement with Takahashi and Maekawa \cite{PRB67_Takahashi2003}, that the measured amplitude of the spin signal in the non-local geometry is strongly reduced for low contact resistances $R_C$ \cite{foot_Rc}. To quantify the effect Popinciuc et al. introduce the so called $R$-parameter that is defined for a 2-dimensional channel by $R=(R_C/R_{sq}) W$, where $R_{sq}$ is the square resistance and $W$ the width of the diffusive channel. \cite{PRB80_Popinciuc2009, foot_2D}  \\
In this article we start by summarizing the dependence of the non-local amplitude on the contact resistance discussed in Ref.~\citenum{PRB80_Popinciuc2009} and then we are going to focus on determining how the Hanle spin precession is influenced by low contact resistances. We discuss that not only the amplitude but also the shape of Hanle precession curves is changed for low values of the $R$-parameter (corresponding to low contact resistances) and simulate Hanle measurements including contact induced relaxation. We quantify the contacts' influence by fitting the data with the standard Hanle formula without taking contact induced effects into account, assuming $R \rightarrow \infty$. Note that fitting with the standard Hanle formula is the common method to analyze experimental spin precession data in almost all published work. The difference of the extracted spin relaxation time $\tau^{fit}$ and diffusion coefficient $D^{fit}$ to the parameters used in the simulations is quantified and we compare these results to data obtained on graphene spin-valve devices where a reduction of $\tau$ was reported for low contact resistances \cite{PRL105_Han2010}. Finally, we point out how to extract correctly the spin transport properties from Hanle precession measurements by excluding spurious background effects. \\

\section{Contact induced spin relaxation}
\begin{figure}
\includegraphics[width=\columnwidth]{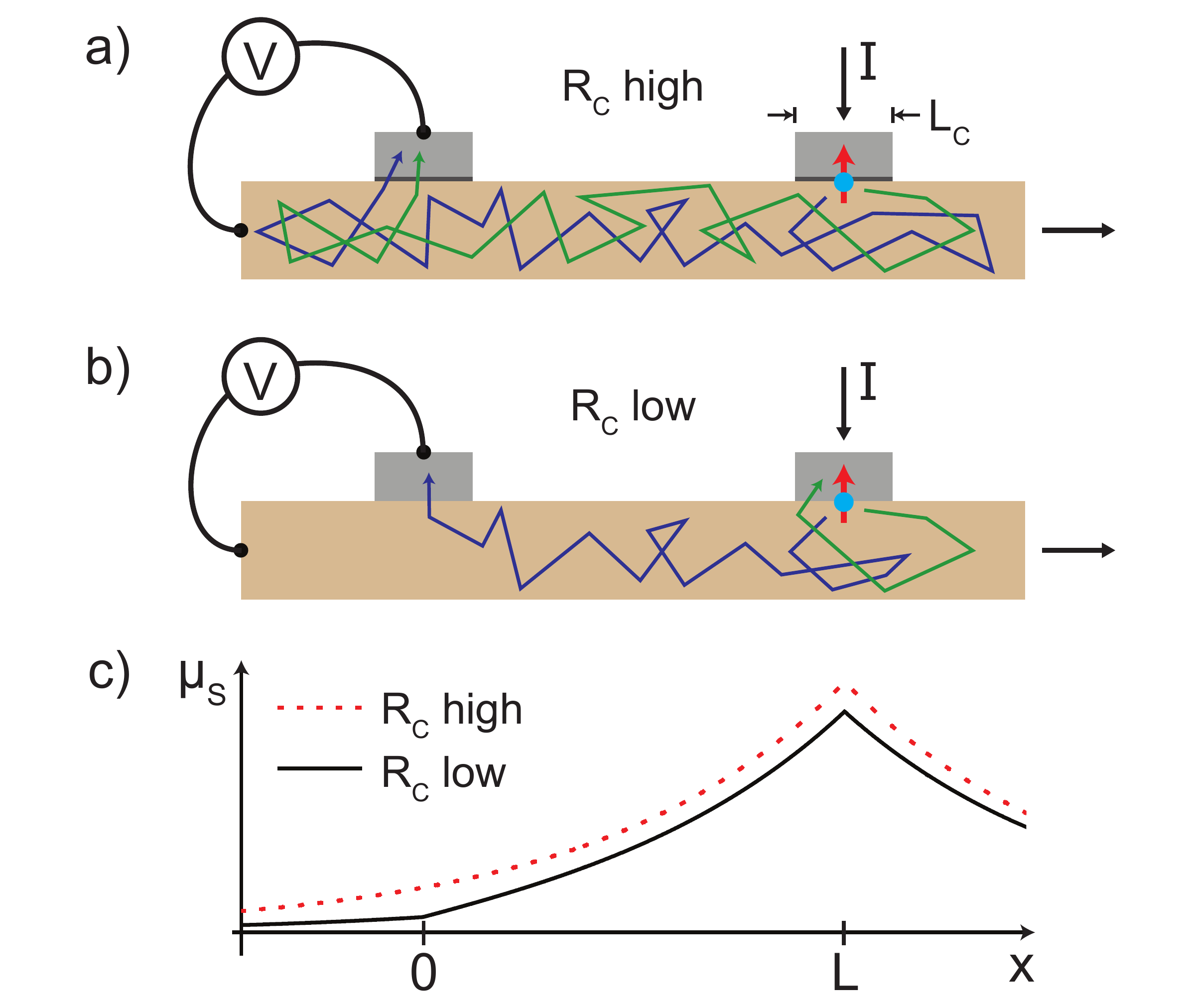} 
\caption{\label{fig:FigI}(Color online) Sketches of spin diffusion through a diffusive channel with a spin injector and detector separated by a distance $L$ in non-local geometry for (a) high and (b) low contact resistances. The width of the contacts in x-direction, $L_C$, is small compared to $L$ or the spin relaxation length $\lambda$. (c) The spin chemical potential $\mu_S$ indicating the spin accumulation below the injector electrode and the exponential decay of the spin signal (red dotted curve). The spin accumulation influenced by the contact induced spin relaxation is denoted by the black solid curve.}
\end{figure}
\begin{figure}
\includegraphics[width=\columnwidth]{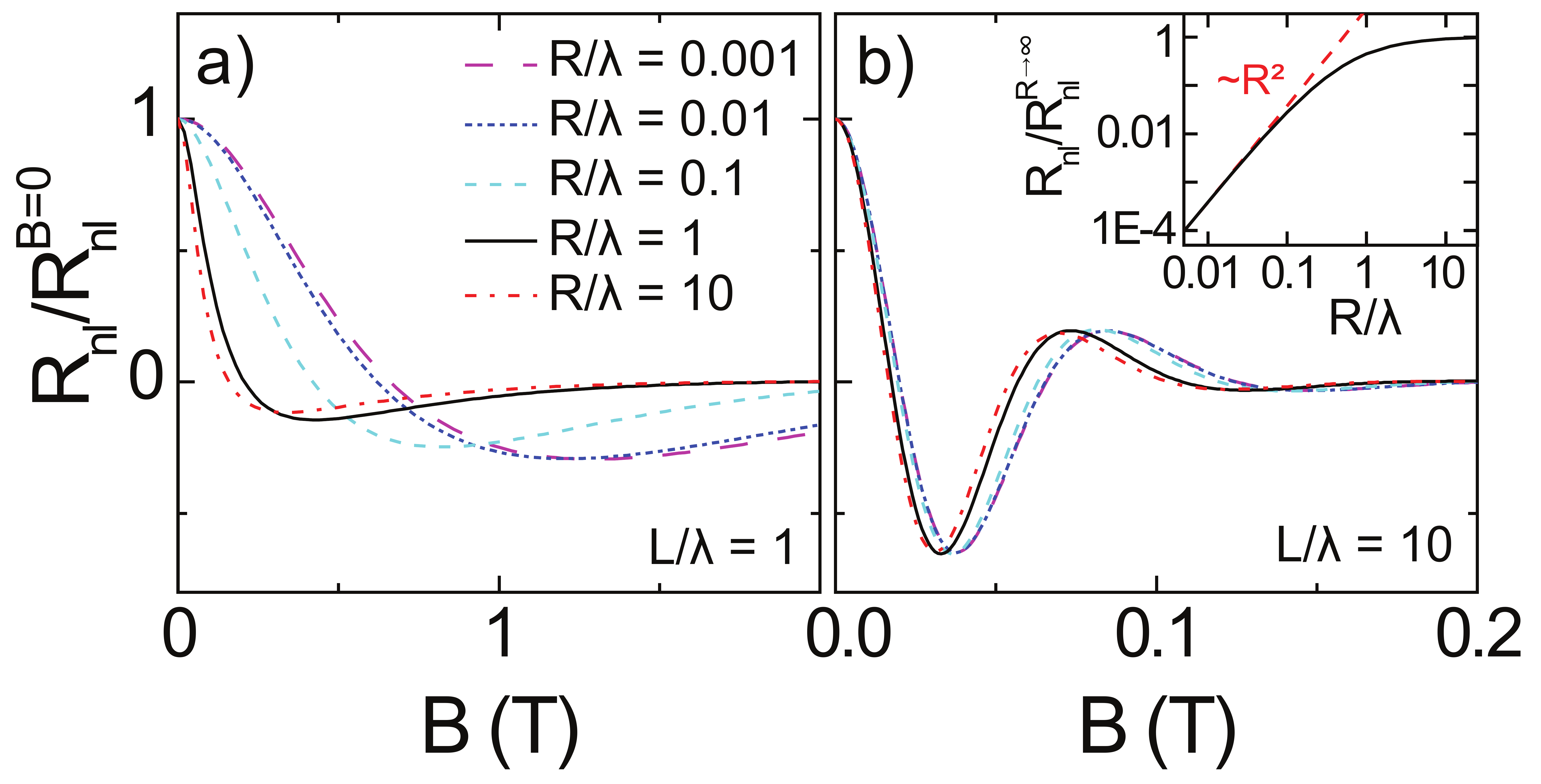} 
\caption{\label{fig:FigII}(Color online) Simulated spin precession curves for different values of $R/\lambda$ with (a) $L/\lambda=1$ and (b) $L/\lambda=10$. For the simulations we use $D=0.01~\mathrm{m^2/s}$, $\tau=100~\mathrm{ps}$, $W=1~\mathrm{\mu m}$ and $R_{sq}=1~\mathrm{k \Omega}$ (representative of graphene devices) with contact resistances between $1$ and $10^4~\mathrm{\Omega}$. The amplitude of the curves is normalized for clarity with $R_{nl}(B=0)$. The inset in panel (b) shows $R_{nl}$ from Eq.~(\ref{eq:RnlAmp}) as a function of $R/\lambda$ for $L/\lambda=1$, normalized by $R_{nl}(R \rightarrow \infty)$ (black solid line) and the asymptote $\propto R^2$ (red dashed line).}
\end{figure}

Fig.~\ref{fig:FigI} (a) presents a sketch of the non-local measurement geometry with an injecting electrode at $x=L$ on the right side, sending a charge current via a resistive barrier \cite{PRB62_Rashba2000} into the channel to the right side end and a detecting electrode on the left side (at $x=0$), measuring the voltage difference between the contact and the left side end. As the electrodes are ferromagnetic, the injecting electrode induces a spin imbalance that accumulates below the electrode and diffuses away from it in both positive and negative x-direction of the channel (red dotted curve in Fig.~\ref{fig:FigI} (c)). %
The second ferromagnetic electrode detects the spins at $x=0$ and the measured voltage is normalized with the injected current to obtain the non-local resistance $R_{nl}$ that is given by \cite{PRB67_Takahashi2003, PRB80_Popinciuc2009, foot_Rc} 
\begin{equation}
R_{nl}=\pm \frac{P^2 R_{sq} \lambda}{2 W} \frac{(2 R/\lambda)^2 \exp(-L/\lambda)}{(1+2 R/\lambda)^2 - \exp(-2L/\lambda)}
\label{eq:RnlAmp}
\end{equation}
The model leading to this result is based on the one-dimensional description of a diffusive channel with an injector and detector on distance $L$ with $P$ the polarization of the contacts and $\lambda=\sqrt{D \tau}$ the spin relaxation length in the channel with the diffusion coefficient $D$. The width of the contacts ($L_C$) is considered to be negligible compared to $L$ and $\lambda$ \cite{PRB80_Popinciuc2009}. Also we assume $1-P^2 \approx 1$ (applicable to graphene devices where $P < 30\%$ \cite{PRL105_Han2010}) and are considering an infinite homogeneous transport channel. The effect of an inhomogeneous transport channel is discussed elsewhere \cite{NL12_Guimaraes2012}. The $R$-parameter is being calculated using the contact resistance of the injector and detector. In case $R$ is not the same for the two electrodes an effective single $R$-parameter can be calculated using $1/R_{eff} \approx (1/R_{1}+1/R_{2})/2$ with the $R$-parameters of the injector and detector $R_1$ and $R_2$ (see Appendix~\ref{sec:Rparameter}). The meaning of the $R$-parameter gets clear when it is normalized with the spin relaxation length $\lambda$. The normalized value corresponds to the ratio of the contact resistance and the spin resistance of the channel $R^s_{ch}=R_{sq} \lambda /W$ so $R/\lambda=R_C/R^s_{ch}$. Hence, $R/\lambda$ describes the ratio of spins diffusing through the channel and relaxing, versus those being reabsorbed by the contact, making it a good measure for the influence of the contacts. \\
Eq.~(\ref{eq:RnlAmp}) shows that the spin signal $R_{nl}$ has a maximum for high contact resistances ($R \rightarrow \infty$) and is reduced for low $R$-values. A significant change is observed for $R/\lambda \leq 1$ (Fig.~\ref{fig:FigII} (b), inset). On the other hand the amplitude of the signal is reduced with increasing $L$ from a maximum at $L=0$. The characteristic length ratio of the system is $L/\lambda$. While the effect on the normalized amplitude ($R_{nl}/R_{nl}^{R \rightarrow \infty}$ with the amplitude without contact induced effects $R_{nl}^{R \rightarrow \infty}$) is smaller for short distances between injector and detector electrode it stays approximately constant for $L/\lambda \geq~1$. Popinciuc et al. discuss in detail the effect of low contact resistances on the measured non-local amplitude but, while included in the model, the effect on the Hanle curve is only discussed qualitatively \cite{PRB80_Popinciuc2009}. In the following we are going to present a quantitative analysis of the influence of low contact resistances on Hanle measurements. We show that the extracted spin transport properties of the transport channel can be limited by the contact induced relaxation and are therefore incorrectly determined when low $R$ measurements are analyzed without considering the influence of the contacts.\\
Fig.~\ref{fig:FigII} (a) shows Hanle precession data that was simulated for different values of $R/\lambda$ with $L/\lambda=1$ using the model system of Fig.~\ref{fig:FigI} (a) described in Ref.~\citenum{PRB80_Popinciuc2009}. Note that the amplitude of the Hanle curves is normalized at $B=0$, which is necessary as the amplitude scales with $(R/\lambda)^2$ for $R/\lambda \ll 1$ and changes by 5 orders of magnitude between $R/\lambda=0.001$ and $R/\lambda=10$. A significant change in the Hanle shape is visible in Fig.~\ref{fig:FigII} (a), pointing to an effective change of the spin transport properties. The strongest change in the shape is seen between $R/\lambda=0.01$ and $R/\lambda=1$ while the curve shape is saturating for both small and large $R/\lambda$ values denoting spin transport limited by the contacts or by the properties of the channel, respectively. Fig.~\ref{fig:FigII} (b) shows a similar dataset for $L/\lambda=10$. We also see a change in the Hanle shape but the effect is much weaker for this larger distance of the injector and detector. Remarkably, in both cases the curves stay in the characteristic Hanle-like shape for all $R/\lambda$. Therefore it is possible to fit the data using the solutions to the Bloch equations \cite{PRB37_Johnson1988, APL81_Jedema2002} that do not take the effect of the low resistive contacts into account (see Appendix~\ref{sec:FitHanle})
\begin{equation}
\frac{d\vec{\mu_S}}{d t} = D\mathbf\nabla^2 \vec{\mu_S} - \frac{\vec{\mu_S}}{\tau} + \vec{\omega_L}\times\vec{\mu_S}.
\label{eq:Bloch}
\end{equation}
Here $\vec{\omega_L}$ is the Larmor frequency $\vec{\omega_L}=g\mu_B/\hbar \ \vec{B}$, with the gyromagnetic factor $g$ (g-factor, $g \approx 2$ for free electrons), the Bohr magneton $\mu_B$ and the magnetic field $\vec{B}$. By fitting simulated data without taking the effects of the contacts into account, we can determine what happens when one fits the data obtained in samples with corresponding $R$- and $L$-values in the standard manner. \cite{foot_Fit}\\
\begin{figure}
\includegraphics[width=\columnwidth]{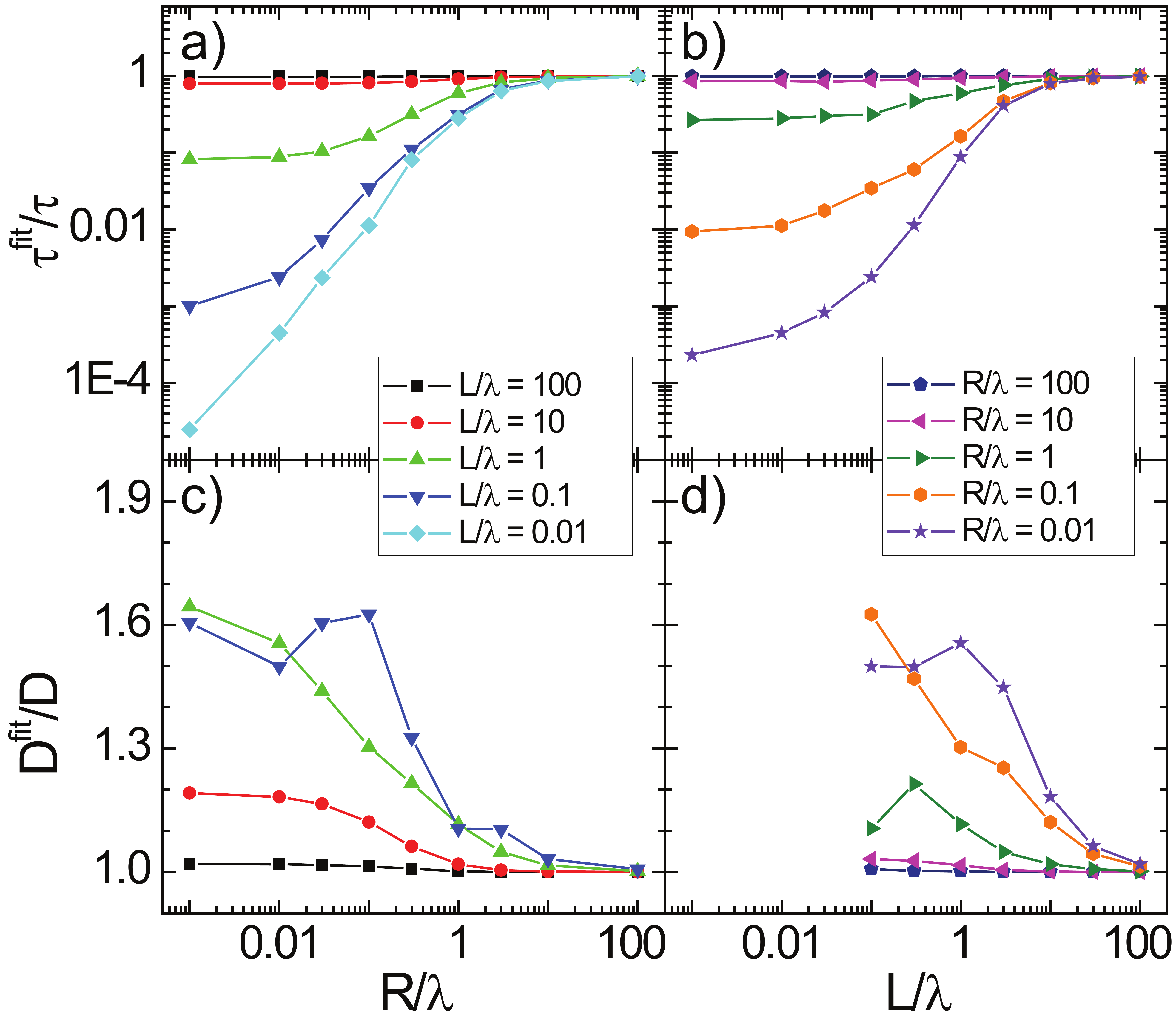} 
\caption{\label{fig:FigIII}(Color online) The change in $\tau^{fit}$ and $D^{fit}$ fitted for different $L/\lambda$ as a function of $R/\lambda$ ((a) and (c)) and for different $R/\lambda$ as a function of $L/\lambda$ ((b) and (d)). For small values of $L/\lambda$ the fits become insensitive to the specific value of the diffusion coefficient, resulting in the non-smooth behavior shown for $L/\lambda = 0.1$ in panel (c). Therefore the data for $L/\lambda < 0.1$ is not shown in panel (c) and (d).}
\end{figure}%
The results from these fits are presented in Fig.~\ref{fig:FigIII}. Note that while the simulations were performed with $D=0.01~\mathrm{m^2/s}$, $\tau=100~\mathrm{ps}$ they do not depend on the specific value of $D$ and $\tau$. Hence, we get the same results for different $D$ and $\tau$ resulting in the same $\lambda=\sqrt{D \tau}$ as the fitting results depend only on the ratios $R/\lambda$ and $L/\lambda$. The graphs show the fitted values $\tau^{fit}$ and $D^{fit}$ normalized by the actual values for the channel $\tau$ and $D$ as a function of $R/\lambda$ for different $L/\lambda$ (Fig.~\ref{fig:FigIII} (a) and (c)) and as a function of $L/\lambda$ with different $R/\lambda$ (Fig.~\ref{fig:FigIII} (b) and (d)). While all values converge for high $R/\lambda$ to the intrinsic values, we see a strong decrease of $\tau^{fit}$ and a moderate increase of $D^{fit}$ for small $R/\lambda$. Looking at Fig.~\ref{fig:FigIII} (a) in more detail we see that the decrease in $\tau^{fit}$ is the strongest the shorter the distance $L$ between injector and detector relative to $\lambda$. We also see that the values saturate for small values of $R/\lambda$ as already perceivable in Fig.~\ref{fig:FigII} (a). In this limit the effect of the contacts is maximized. $\tau^{fit}$ shows changes of nearly up to five orders of magnitude which means that in a measurement with parameters of $R/\lambda=0.001$ and $L/\lambda=0.01$ we would underestimate $\tau$ by a factor of $5 \times 10^4$.\\
The length dependence of the effect is more clearly presented in Fig.~\ref{fig:FigIII} (b) where the $\tau^{fit}$ data is plotted as a function of $L/\lambda$ for different $R/\lambda$. Here we see that while the decrease of $\tau^{fit}$ is stronger for shorter distances the effect gets negligible for $L/\lambda\geq10$. That means that contact induced effects can be circumvented by measuring on a longer distance. This is only limited by the reduced measured amplitude for longer distances $L$ (see Eq.~(\ref{eq:RnlAmp})). \\
Fig.~\ref{fig:FigIII} (c) and (d) show the same plots for $D^{fit}$. Also here we see the strongest effect for small $R/\lambda$ and $L/\lambda$ and no significant change for $R/\lambda=100$ or $L/\lambda=100$. On the other hand, the values for $D^{fit}$ show a much weaker change than the values of $\tau^{fit}$ and the change is directed in the opposite direction than the change of $\tau^{fit}$. Similar to $\tau^{fit}$, the $D^{fit}$ values also seem to saturate for small $R/\lambda$ and the changes are less than a factor of $2$. \\
While most curves presented in Fig.~\ref{fig:FigIII} have a smooth shape and a continuous change with $L/\lambda$ and $R/\lambda$, the data for $D^{fit}$ shows for values of $L/\lambda \leq 0.1$ combined with values of $R/\lambda \leq 1$ a non-smooth behavior. This is related to the fact that the diffusion in the channel gets for small $L$ dominated by the contact induced effects for short distances and low contact resistances and the shape of the Hanle curves gets strongly influenced. The spin accumulation has no significant decay between the injector and the detector electrode so the system becomes similar to 3-terminal Hanle precession \cite{N462_Dash2009}. As a result, the fits become insensitive to the specific value of $D$ and one can only determine $\tau$ \cite{c_Zomer2012} (see Appendix~\ref{sec:FitHanle}). Therefore we omitted the data for $D^{fit}$ for $L/\lambda < 0.1$ in Fig.~\ref{fig:FigIII} (c) and (d). \\
Note that in the limit $L/\lambda \ll 1$ and $R/\lambda < 10$ the values for $\tau^{fit}$ saturate as they are dominated by the contact induced effects and can be described by a basic formula related to the back diffusion of the spins into the contact (see Appendix~\ref{sec:explainTau}).
\section{Discussion}
Fig.~\ref{fig:FigIII} shows clear trends for $\tau^{fit}/\tau$ and $D^{fit}/D$ as a function of $R/\lambda$ and $L/\lambda$. We are going to discuss in the following how to understand the physics behind the presented results. The sketch in Fig.~\ref{fig:FigI} (a) presents the spin injection and detection for high contact resistances, e.g. due to tunnel barriers between the channel and the contacts. Here the spin diffusion in the channel remains undisturbed and the injected spins diffuse freely through the channel before being detected by the spin sensitive detector. In this way, measurements detect the intrinsic spin transport properties of the channel and the simple exponential decay of the spin signal (red dotted curve in Fig.~\ref{fig:FigI} (c)) is obtained.\\
In the case of low contact resistances the spin transport is influenced both at the injector and at the detector electrodes. When diffusing through the channel the low resistive detector has a high probability of detecting the spins as soon as they are near the contact as it acts as a spin sink (Fig.~\ref{fig:FigI} (b)). Therefore the effective traveling time is reduced and the measured diffusion coefficient enhanced as $D=L^2/2\tau_D$ where $\tau_D$ is the diffusion time for the length $L$. At the same time the proximity to the low resistive contacts also causes spins to relax, which reduces the relaxation time. The extra relaxation is depicted by the kink at the detector in the black solid curve in Fig.~\ref{fig:FigI} (c), describing the decay of the in general reduced spin accumulation in the system.\\
Fig.~\ref{fig:FigIII} (b) and (d) show a reduction of the contact induced effects for larger $L/\lambda$. This can be easily understood by the fact that for a longer distance between the electrodes the ratio of the time the spins stay in the channel compared to in close proximity to the contacts grows resulting in relatively less influence of the contacts on the spin transport.\\
\begin{figure}
\includegraphics[width=\columnwidth]{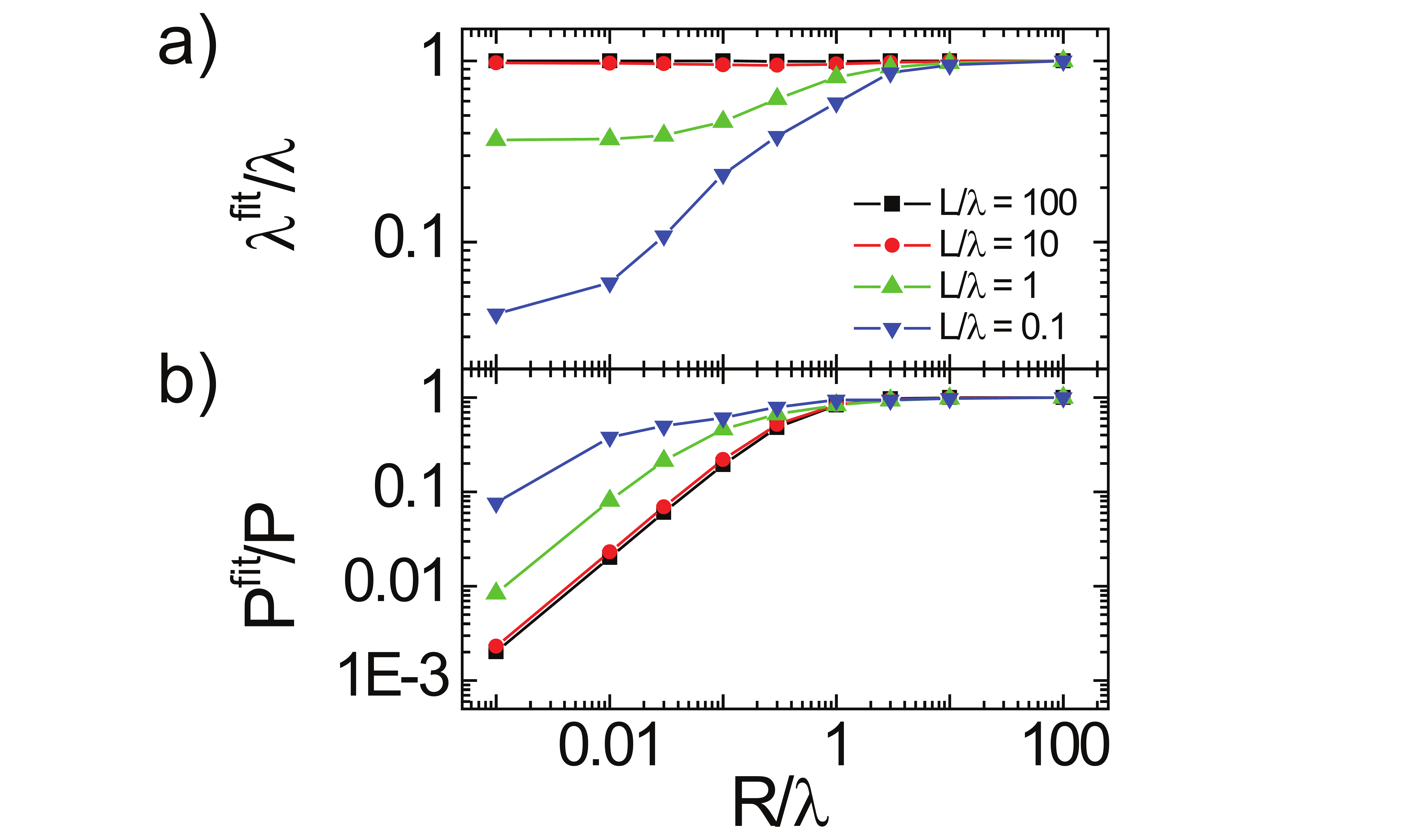} 
\caption{\label{fig:FigIV}(Color online) (a) The change in $\lambda^{fit}/\lambda$ calculated using $\tau^{fit}$ and $D^{fit}$ from Fig.~\ref{fig:FigIII}. (b) The effective polarization $P^{fit}$ normalized with the actual polarization $P$. The values are plotted as a function of $R/\lambda$ for different $L/\lambda$. The values for $L/\lambda=100$ and $10$ overlap in both panels. }
\end{figure}%
Fig.~\ref{fig:FigIV} (a) shows the spin relaxation length $\lambda^{fit}$ resulting from the fitting results presented in Fig.~\ref{fig:FigIII} for $L/\lambda \geq 0.1$. The shape of the $\lambda^{fit}$-curve is comparable to the behavior of $\tau^{fit}$. This is due to the fact that $\lambda^{fit}$ is mainly influenced by the spin relaxation time $\tau^{fit}$ with a change of up to a factor $1000$ (for $L/\lambda \geq 0.1$). As $D^{fit}$ shows only a change of less than a factor $2$ we get a maximum reduction of $\lambda^{fit}$ by a factor $25$. \\
This $\lambda^{fit}$ value would be used in the analysis of a measurement to calculate the polarization $P$. If we take the amplitude simulated with Eq.~(\ref{eq:RnlAmp}) (inset, Fig.~\ref{fig:FigII} (b)) and assume spin transport without contact induced spin relaxation (Eq.~(\ref{eq:RnlAmp}) for $R \rightarrow \infty$) we extract the effective polarization $P^{fit}$ with $R_{nl}(\lambda^{fit}, P^{fit}, R \rightarrow \infty)=R_{nl}(\lambda, P, R)$. The resulting value is up to $500$ times reduced for small values of $R/\lambda$ compared to the real $P$ value (Fig.~\ref{fig:FigIV} (b)). Note that the largest change in $P^{fit}$ compared with $P$ is observed for long distances. \\
After discussing the effects observed in the simulations let us have a look on measurements on real devices using graphene as the transport channel. In spin transport samples in graphene it is difficult to produce high resistive contacts and to control the quality of the contact-graphene interface. So a data set with similar quality samples with only a change of the contact resistance is difficult to produce. On the other hand in a single device the quality of the contacts is most of the time comparable. Therefore it is relatively easy to check the length dependence of the spin transport properties in this kind of system, assuming similar $R$-parameters for all electrodes. Two sets of data obtained on two different graphene devices with three different injector-detector distances are presented in the work by Wojtaszek et al. in Fig.~4~(a) and Fig.~5~(a) of the supplementary information of Ref.~\citenum{c_Wojtaszek2012}. In both cases a minor increase of $D$ is reported when measuring on a shorter distance and in the first case also a minor decrease of $\tau$, pointing to weak but apparent contact induced relaxation \cite{foot_data2}. With $R \geq 3~\mathrm{\mu m}$ and $\lambda \approx 5~\mathrm{\mu m}$ the measurements were also performed in a regime where one would expect this kind of weak contact induced effects as $L/\lambda \approx R/\lambda \approx 1$ (see Fig.~\ref{fig:FigIII}) \cite{c_Wojtaszek2012}. \\
Han et al. present a study of the dependence of the spin transport properties on the quality of the resistive barrier between the graphene channel and the contacts in Ref.~\citenum{PRL105_Han2010}. They show that between tunneling injection of spins and the injection with transparent contacts the measured spin relaxation time decreases while the diffusion coefficient is increased in agreement with our simulations' results. On the other hand the results for a ``pinhole'' barrier with intermediate resistance present an intermediate spin relaxation time but also a reduced diffusion coefficient. While the spin relaxation time fits into the expectations for an intermediate contact resistance, the reduced diffusion coefficient cannot be explained by the contact resistance but has to be related to a lower quality sample or other effects.\\
Our model also points to the fact that the recent reported differences between the results for the spin relaxation length, based on the analysis of 4-terminal non-local Hanle precession measurements \cite{PRL105_Han2010} and based on the analysis of the magnitude of spin-valve measurements in local 2-terminal geometry with very high contact resistances ($R_C > 1~\mathrm{M \Omega}$) \cite{NPaoP_Dlubak2012} cannot be explained by contact induced relaxation. If one would measure with the configuration of Han et al.\cite{PRL105_Han2010} with $L=5.5~\mathrm{\mu m}$ and $R\approx 200~\mathrm{\mu m}$ a material with a spin relaxation length of $\sim 100~\mathrm{\mu m}$ and a spin relaxation time of $\sim 100~\mathrm{ns}$ as reported in Ref.~\citenum{NPaoP_Dlubak2012} one would only see a reduction of the fitted spin relaxation time by a factor of $\tau^{fit}/\tau \approx 1/3$ (see Fig.~\ref{fig:FigIII} (a)) leading to a reduced spin relaxation length of $\lambda^{fit}/\lambda \approx 1/2$ (see Fig.~\ref{fig:FigIV} (a)) as one would have $L/\lambda \approx 0.05$ and $R/\lambda\approx 2$. Therefore the standard Hanle analysis would yield $\lambda^{fit}\approx50~\mathrm{\mu m}$ and $\tau^{fit}\approx30~\mathrm{ns}$ but Han et al. report $\lambda^{fit} \approx 2.5~\mathrm{\mu m}$ and $\tau^{fit}\approx0.5~\mathrm{ns}$ \cite{PRL105_Han2010}. With $\lambda^{fit} \approx 2.5~\mathrm{\mu m}$ Han et al. are in the regime of negligible contact induced relaxation with $L/\lambda \approx 2$ and $R/\lambda \approx 80$, so the difference in the measured $\lambda$ is not based on contact induced relaxation but has to be related to other effects. Even for a spin relaxation length of $\lambda=20~\mathrm{\mu m}$ it would be $L/\lambda \approx 0.25$ and $R/\lambda \approx 10$ for the system of Han et al. and they would be able to measure this $\lambda$ without significant influence of the contacts (see Fig.~\ref{fig:FigIV} (a)). Such strong differences of the spin signal magnitude between non-local and local configuration as between Refs.~\citenum{PRL105_Han2010} and \citenum{NPaoP_Dlubak2012} have also been observed in traditional semiconductors like silicon in the non-local \cite{APE4_Suzuki2011} and 3-terminal \cite{N462_Dash2009} configuration.

\section{Guidelines for a good and reliable Hanle fit}
\begin{figure}
\includegraphics[width=\columnwidth]{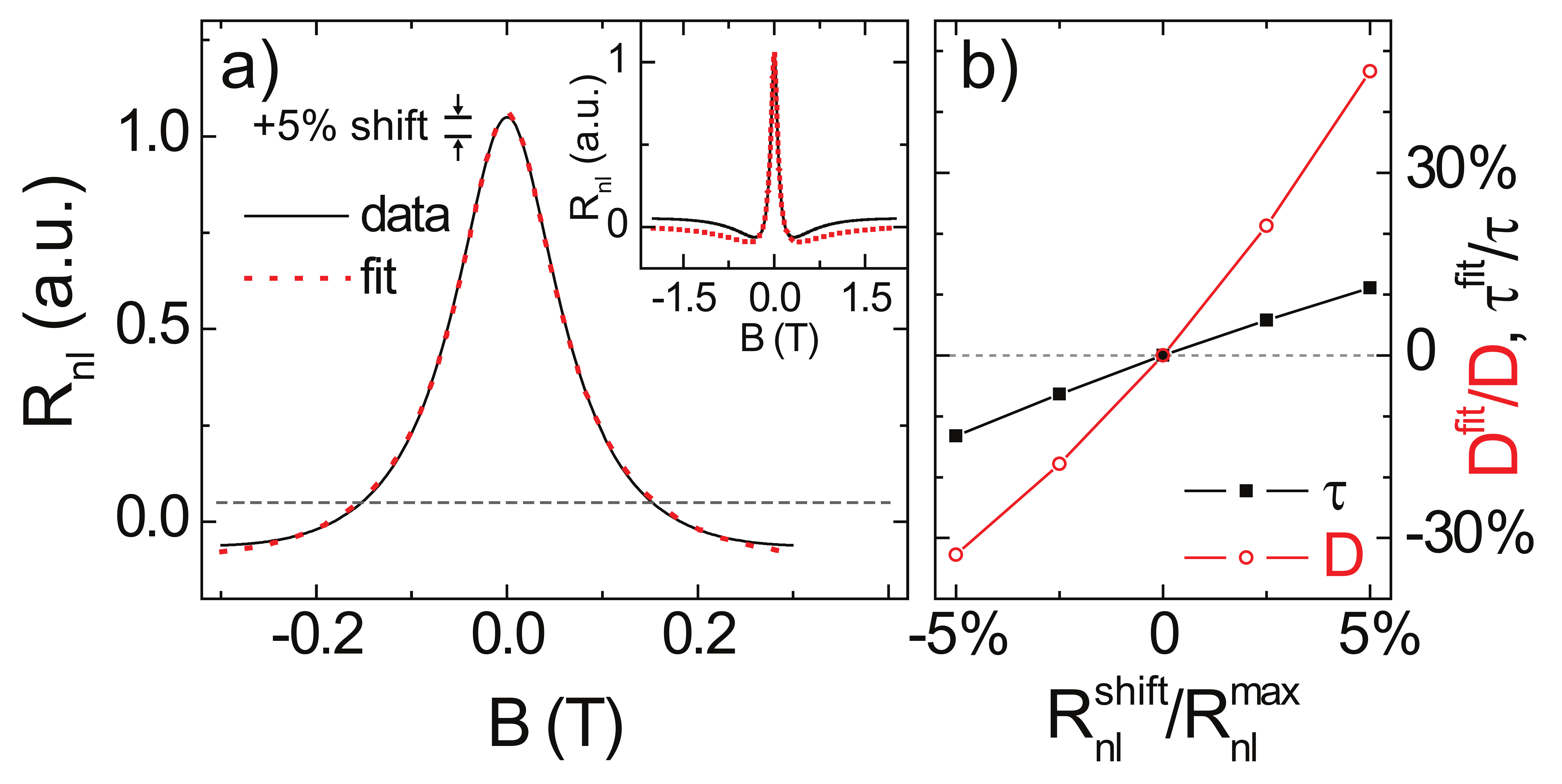} 
\caption{\label{fig:FigV}(Color online) (a) The influence of a baseline shift shown by means of a Hanle spin precession curve, shifted $+5\%$ of the precession amplitude upwards (black solid curve) and a fit assuming no shift (red dotted curve). The baseline of the fit is therefore at $R_{nl}=0$ while the baseline of the data is denoted by the black dashed line. The same Hanle curves on a larger $B$-field range are shown in the inset. A clear difference is visible for $|B| > 0.3~\mathrm{T}$. (b) The change of the diffusion coefficient and the spin relaxation time resulting from data with a baseline shift and fits assuming no baseline shift. The presented data was simulated using $D=0.01~\mathrm{m^2/s}$, $\tau=100~\mathrm{ps}$ and $L=1~\mathrm{\mu m}$.}
\end{figure}%
In this paper we discuss how Hanle measurements are influenced by contact induced relaxation that can lead to incorrectly determined spin transport properties of the channel. Independently from that, the fitting procedure can also give incorrect results for the spin transport properties when performed incorrectly. In this section we are therefore commenting on typical pitfalls in analyzing Hanle precession data. \\
The fit to a Hanle curve is unambiguous if performed in the right way. Fig.~\ref{fig:FigV} (a) illustrates how a fit can still give wrong results on the example of a Hanle precession fit when assuming a wrong background resistance. The background resistance is represented by the $R_{nl}(B \rightarrow \infty)$ value and is the fitting baseline. Fig.~\ref{fig:FigV} (a) shows a fit to the central peak of a Hanle curve (without contact induced effects) with a baseline shifted by $+5 \%$ of the amplitude. The fit results in an increase of $\tau$ by $>10\%$ and of $D$ by $>45\%$ and therefore a misestimation of $\lambda$ by more than $25 \%$ compared to the values used to simulate the data. However, when fitting the curve with these values the fit presents itself faulty when including the high field tails of the curve as shown in the inset of Fig.~\ref{fig:FigV} (a). Fitting to high $B$-field values gives therefore a good indication of the quality of the fit. However, this identification of a bad fit can be partly masked by data noise in combination with anisotropic magneto resistance effects or the out-of-plane tilting of the magnetization of the ferromagnetic electrodes at high field values, adding an additional background resistance \cite{PRL101_Tombros2008}. Another indication of a good fit is the fitted curve reproducing the ``shoulders'' of the measured curve, where $R_{nl}$ has a minimum (for parallel alignment of the injector and detector electrode). This is obviously not accomplised in the presented case (Fig.~\ref{fig:FigV} (a), inset). The larger the ratio $L/\lambda$ the more pronounced are the shoulders, so measuring on a longer distance enhances the reliability of the fit.\\
While measuring to high magnetic field values to determine the background resistance is in any case advisable, there is a way to avoid such spurious background effects in a fit. Measuring the spin precession both for parallel and for antiparallel orientation of the electrodes, and subtracting the signals from each other removes most spurious (not spin related) background effects as done in several recent works \cite{NP8_Vera-Marun2012, NL12_Guimaraes2012, c_Wojtaszek2012, c_Zomer2012}. By taking the mean of the parallel and the antiparallel measurements, one can also extract the $B$-field dependent background resistance. Finally, a minor error in a fit can also occur if the magnetic field values are not properly calibrated. The effect of a correction factor for the magnetic field value is the same as the effect of a changed g-factor as $\omega_L \propto B$ and $\omega_L \propto g$. A wrong $B$-field calibration is therefore linearly passed on to $\tau$ and $1/D$ \cite{c_Maassen2012}.
\section{Conclusions}
We discuss the effect of low resistance contact induced spin relaxation on Hanle precession data and quantify the misinterpretation of spin transport properties in a transport channel that can arise from this effect. As fitting Hanle curves is a common way to extract spin transport properties we use the model presented in Ref.~\citenum{PRB80_Popinciuc2009} to simulate Hanle measurements and fit the data using the standard formula, neglecting the contact induced effects. The observed rescaling of the spin relaxation time and the diffusion coefficient only depend on the ratios $R/\lambda$ and $L/\lambda$ and the fitting results show that a strongly decreased $\tau^{fit}$ by up to nearly five orders of magnitude and a moderately increased $D^{fit}$ by less than a factor of $2$ can be observed for small $R/\lambda$ and $L/\lambda$. On the other hand large values for both $R/\lambda$ or $L/\lambda$ show a convergence of $\tau^{fit}$ and $D^{fit}$ on the undisturbed values $\tau$ and $D$, independent on $L$ or $R$, respectively. This shows that the spin relaxation induced by the contacts can in principle be avoided when measuring on a longer distance. We then discuss how these values of $\tau^{fit}$ and $D^{fit}$ lead to a wrong estimate of the contact polarization before comparing our results for $\tau^{fit}$ and $D^{fit}$ qualitatively with measurements on graphene in the literature. The modeled effect of the contacts on spin transport only depends on the resistance of the barrier and not on the type of barrier. Hence, although most contact interfaces used in the non-local geometry to study spin transport in graphene are not truly in the tunneling regime, we can conclude that with the the resistance of the commonly used barriers the effect of back diffusion into the contacts on the spin transport is only minor and the spin transport properties are mainly limited by other effects. \cite{N448_Tombros2007,PRB80_Jozsa2009,PRB80_Popinciuc2009,PRL105_Han2010, PRB83_Maassen2011,PRL107_Han2011, PRL107_Yang2011, NP8_Vera-Marun2012, NL12_Guimaraes2012, c_Wojtaszek2012, c_Zomer2012} While explicitly discussing the effect of low resistive contacts on the non-local geometry, similar effects also play a role for local measurements \cite{PRB82_Jaffres2010}.\\% (eq (11) and (12)) 
We also briefly discussed the guidelines for a good and reliable Hanle fit as an incorrectly performed fit can also lead to misinterpretations of the spin transport properties of a diffusive channel while a correct fit leads to unambiguous results.

\begin{acknowledgments}
We would like to thank M. Wojtaszek for useful discussions. This work was financed by the Zernike Institute for Advanced Materials and the Foundation for Fundamental Research on Matter (FOM).
\end{acknowledgments}

\appendix 
\section{\label{sec:Rparameter}The $R$-parameter for dissimilar contacts}
The discussion in the main text focused on the symmetric case when the injector and the detector contacts have equal $R$-parameters. Here we address the general case of dissimilar injector and detector contacts and demonstrate an equivalence that allows us to map this general case to the more symmetric one presented above.

The general expression for the non-local resistance $R_{nl}$ given by \cite{PRB67_Takahashi2003}
%\begin{widetext}  
\begin{equation}
\begin{split}
R_{nl}=&\pm \frac{R_{sq} \lambda}{2 W} \exp{(-L/\lambda)} \prod_{i=1}^{2} \left( \frac{P \dfrac{2 R_i}{\lambda}}{1-P^2} \right) \\
& \times \left[ \prod_{i=1}^{2} \left( 1 + \frac{ \dfrac{2 R_i}{\lambda}}{1-P^2} \right) - \exp{(-2L/\lambda)} \right]^{-1}
\end{split}	
\label{eq:RnlAmpFull}
\end{equation}
%\end{widetext} 
where $R_{1,2}$ correspond to the $R$-parameters of the injector and detector contacts, and the rest of the parameters are the same as those presented in the discussion of Eq.~(\ref{eq:RnlAmp}) \cite{foot_Rc}.

\begin{figure}[tbp]
\includegraphics[width=\columnwidth]{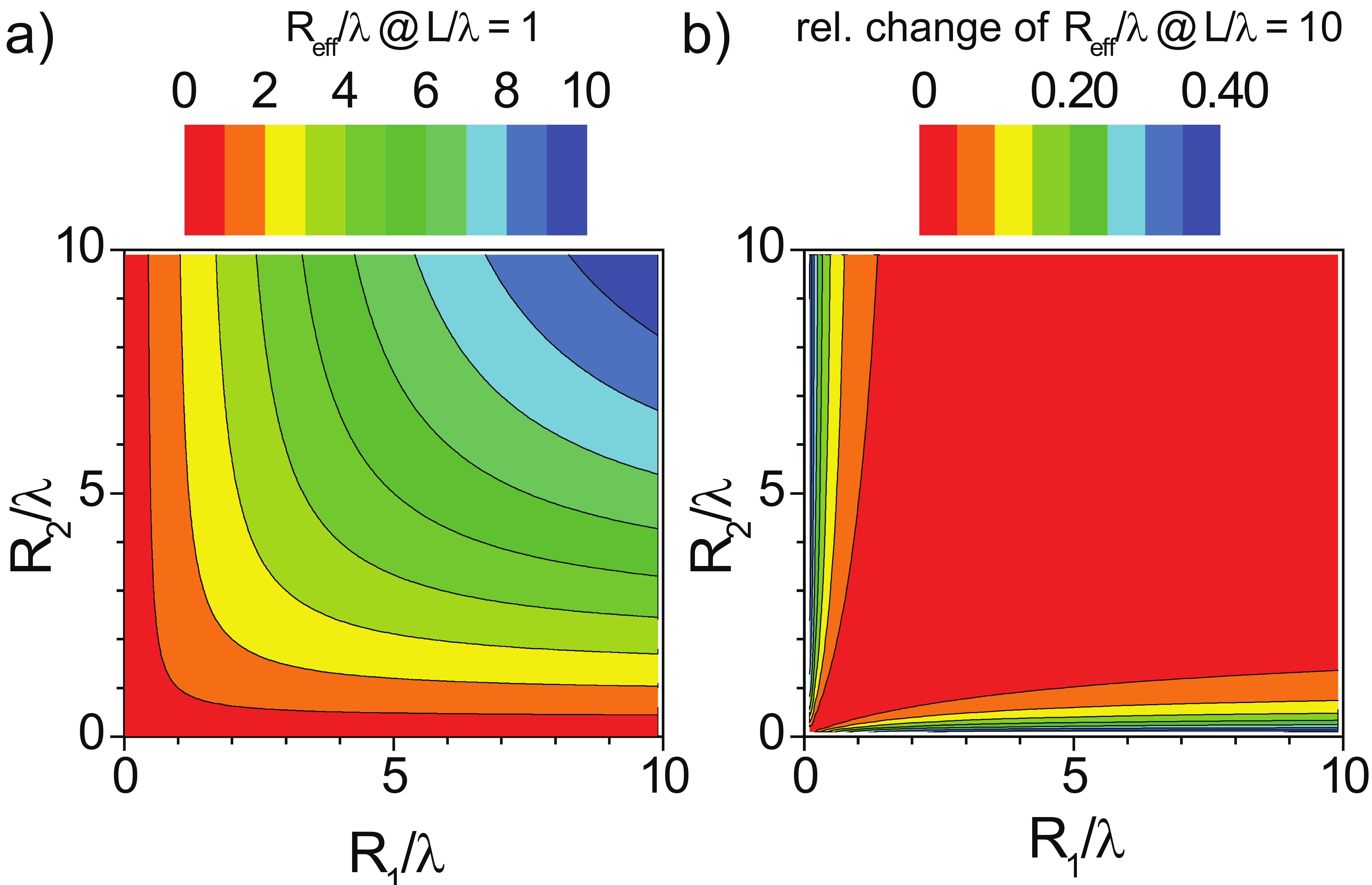} 
\caption{\label{fig:FigAI}(Color online) Mapping the problem of dissimilar contacts $R_1 \neq R_2$ into the simpler one of identical contacts with a common $R_{eff}$. (a) Two-dimensional map of equivalent $R_{eff}/\lambda$ as a function of $R_i/\lambda$ of the contacts for $L/\lambda = 1$ from Eq.~(\ref{eq:RnlAmpMap}). (b) Normalized deviation of $R_{eff}/\lambda$ obtained from Eq.~(\ref{eq:RnlAmpMap2}) relative to the exact result from Eq.~(\ref{eq:RnlAmpMap}) in the limit of $L/\lambda \gg 1$. The values are normalized using ${(R_{eff}(L/\lambda=10)}-R_{eff}(L/\lambda=0))/(R_{eff}(L/\lambda=10)$.}
\end{figure}

This equation can be simplified by realizing that for highly spin polarized contacts ($P \approx~1$) there is no contact induced spin relaxation, even for low resistance contacts. Therefore if we consider $1-P^2 \approx 1$ we obtain 
\begin{equation}
R_{nl}=\pm \frac{ P^2 R_{sq} \lambda}{2 W} \frac{ \left( \dfrac{2 R_1}{\lambda} \right) \left( \dfrac{2 R_2}{\lambda} \right) \exp{(-L/\lambda)} } { \left(1 + \dfrac{2 R_1}{\lambda}\right) \left(1 + \dfrac{2 R_2}{\lambda}\right) - \exp({-2L/\lambda}) }
\label{eq:RnlAmpFull2}
\end{equation}
which has a similar structure as Eq.~(\ref{eq:RnlAmp}) \cite{PRB80_Popinciuc2009}. Following simple algebra, we can equate both equations and solve for the $R$-parameter of Eq.~(\ref{eq:RnlAmp}), which can be understood as an effective $R$-parameter $R_{eff}(R_1,R_2,L,\lambda)$ given by,
\begin{widetext} 
\begin{equation}
\frac{2 R_{eff}}{\lambda} =
\frac{
\left( \frac{2 R_1}{\lambda} \frac{2 R_2}{\lambda} \right) + 
	\sqrt{
	\left( \frac{2 R_1}{\lambda} \frac{2 R_2}{\lambda} \right)^2
	- \left( 1+ \frac{2 R_1}{\lambda} + \frac{2 R_2}{\lambda} - e^{-2L/\lambda} \right)
	\left( \frac{2 R_1}{\lambda} \frac{2 R_2}{\lambda} \right)
	(e^{-2L/\lambda}-1)
	}
}
{
	1 + \frac{2 R_1}{\lambda} + \frac{2 R_2}{\lambda} -
	e^{-2L/\lambda}
}
\label{eq:RnlAmpMap}
\end{equation}
\end{widetext}
allowing us to map the case of dissimilar contacts into the symmetric case of equal contacts with $R_{1,2} = R_{eff}$. One example of such a mapping is shown in Fig.~\ref{fig:FigAI} (a) for the representative case of $L/\lambda = 1$ and $R_i/\lambda = 0.1\text{--}10$. We observe that when $R_1 \neq R_2$ then $R_{eff} \approx \min{(R_1,R_2)}$ and also the trivial case of $R_1=R_2=R_{eff}$.

The exact mapping depends on $L$ and on $\lambda$, which requires careful application to analyze experimental data. We remark that this issue is absent for the case of $2L/\lambda \approx 0$, where Eq.~(\ref{eq:RnlAmpMap}) reduces to the simple form,
\begin{equation}
\frac{1}{R_{eff}} = \frac{1}{2}\left( \frac{1}{R_1} + \frac{1}{R_2} \right)
\label{eq:RnlAmpMap2}
\end{equation}
equivalent to a 3-terminal measurement where both contacts are in a parallel configuration.

Although Eq.~(\ref{eq:RnlAmpMap2}) is strictly speaking valid only when both contacts are closely spaced, we have observed that it offers a reasonable approximation even at finite separation $L$ between the contacts. In Fig.~\ref{fig:FigAI} (b) we compare the resulting $R_{eff}/\lambda$ for the extreme case of large separation ($2L/\lambda \gg 1$) from Eq.~(\ref{eq:RnlAmpMap}), to the value obtained from the simpler Eq.~(\ref{eq:RnlAmpMap2}). Surprisingly, in the experimentally relevant range of intermediate conductivity mismatch $R_i/\lambda = 0.3\text{--}10$, Eq.~(\ref{eq:RnlAmpMap2}) deviates from the exact result at infinite separation only by less than 20\%. For a strong conductivity mismatch ($R_i/\lambda \leq 0.1$) one should apply the exact result of Eq.~(\ref{eq:RnlAmpMap}).

\section{\label{sec:FitHanle}Fitting simulated Hanle curves}
\begin{figure}
\includegraphics[width=\columnwidth]{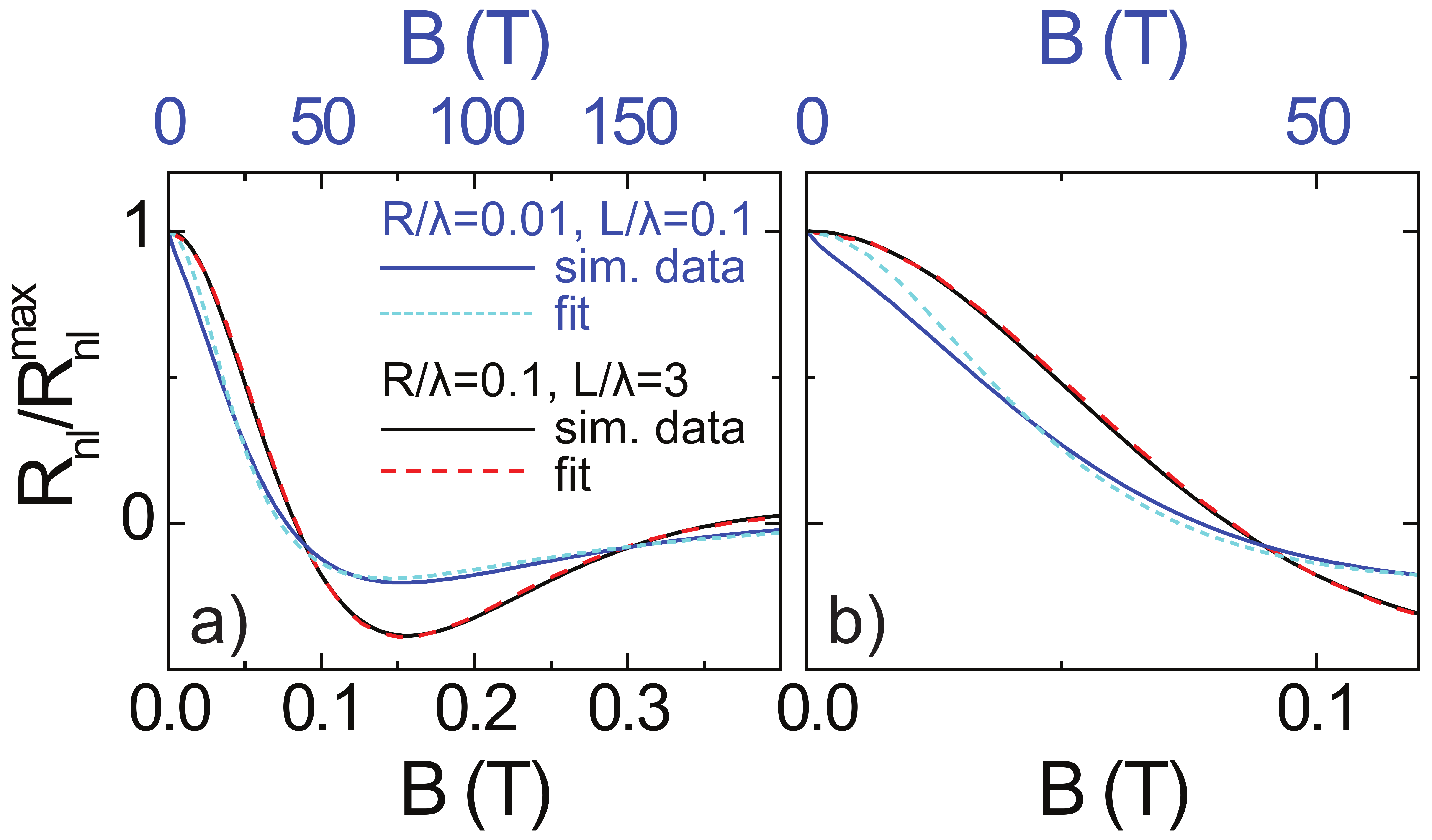} 
\caption{\label{fig:FigAII}(Color online) Two sets of simulated data with the corresponding fits assuming an amplitude of $1$ and a baseline at $0$. Both sets were simulated for $D=0.01~\mathrm{m^2/s}$, $\tau=100~\mathrm{ps}$ and the values for $L/\lambda$ and $R/\lambda$ shown in the legend. (a) shows the full Hanle curve, (b) zooms in on the part close to $B=0$. The $B$-scale for the curve with the red dashed fit is on the bottom, the scale for the curve with the light blue dotted fit on top.}
\end{figure}%
The research presented in this paper is based on the following concept: Hanle precession curves are simulated following the model presented in Ref.~\citenum{PRB80_Popinciuc2009} including contact induced spin relaxation and are fitted neglecting the contact induced effects. Fig.~\ref{fig:FigAII} shows how well the simulated data can be fitted with a Hanle curve for different values of $R/\lambda$ and $L/\lambda$. The curve for $R/\lambda=0.1$ and $L/\lambda=3$ shows that even for small $R/\lambda$-values (corresponding to a contact resistance of $R=100~\Omega$ when $R_{sq}=1~\mathrm{k \Omega}$ and $W=1~\mathrm{\mu m}$) we get an excellent fit (although with a reduced $\tau^{fit}$ and increased $D^{fit}$). On the other hand the curve and fit for the combination $R/\lambda=0.01$ and $L/\lambda=0.1$ points out that the fit is not describing the curve properly for very small values of the two parameters. This is especially well visible close to $B=0$ (Fig.~\ref{fig:FigAII} (b)) where due to the strong contact induced relaxation we observe a distinct drop of $R_{nl}$ which the fit cannot describe. In Fig.~\ref{fig:FigIII} (c) and (d) of the main text it is visible for which sets of parameters the fits do not describe the Hanle curves well, as those are the points that do not show a smooth line shape when plotting $D^{fit}$ as a function of $R/\lambda$ or $L/\lambda$.
\section{\label{sec:explainTau}$\tau^{fit}$ in the limit $L/\lambda \ll 1$}
We can obtain $\tau^{fit}$ by performing a standard Hanle fit on simulated data that includes contact induced relaxation. Here we show that we can approximate the value of $\tau^{fit}$ for small $L$ using an easy reasoning.\\
In the limit $L/\lambda \ll 1$ our system resembles the 3-terminal Hanle geometry \cite{N462_Dash2009} as we have two contacts connected to approximately the same point of the transport channel with one of the contacts injecting spins and the other detecting them. At the same time there is the transport channel pointing in two directions away from the injection point. Therefore we get for the spin resistance $1/R_{spin}^\ast=1/R^s_{ch}+1/R_C$. If we now take the ratio of the spin resistance including the contact resistance ($R_{spin}^\ast$) and $R_{spin}=R^s_{ch}$ ($R_{spin}^\ast$ for $R_C \rightarrow \infty$) we get: 
\begin{equation}
\frac{R_{spin}^\ast}{R_{spin}}=\frac{1/R_{spin}}{1/R_{spin}^\ast}=\frac{1/R^s_{ch}}{1/R^s_{ch}+1/R_C}=\frac{R/\lambda}{1+R/\lambda}
\label{eq:RspinoverRspin}
\end{equation}
The spin resistance is proportional to the non-local signal ($R_{spin} \propto R_{nl}$) and for $L=0$ and $R \rightarrow \infty$ the non-local signal is proportional to the spin relaxation length $R_{nl} \propto \lambda$ (see Eq.~(\ref{eq:RnlAmp})). Therefore we get: 
\begin{equation}
\frac{R_{spin}^\ast}{R_{spin}}=\frac{\lambda^{fit}}{\lambda} \approx \sqrt{\frac{\tau^{fit}}{\tau}}
\label{eq:RspinTau}
\end{equation} 
We can use here for both $R_{spin}^\ast$ and $R_{spin}$ the relation $R_{nl} \propto \lambda$. This is obviously valid for $R_{spin}$ and for $R_{spin}^\ast \propto \lambda^{fit}$ we have to keep in mind that $\lambda^{fit}$ is obtained assuming $R \rightarrow \infty$ so we have to assume this also here in this analysis. \\
The relation between the ratio of the spin relaxation lengths and the ratio of the spin relaxation times is valid as $D^{fit}/D \approx 1$.\\
Hence, we get the result: 
\begin{equation}
\frac{\tau^{fit}}{\tau}\approx\left(\frac{R/\lambda}{1+R/\lambda}\right)^2.
\label{eq:TauofR}
\end{equation} 
In the limit $L/\lambda \ll 1$ we therefore expect $\tau^{fit}/\tau=(0.98, 0.83, 0.25, 8.3\times10^{-3}, 9.8\times10^{-5})$ for $R/\lambda=(100,10,1,0.1,0.01)$ in good agreement with the values in Fig.~\ref{fig:FigIII} (b) in the limit $L/\lambda \ll 1$.\\
%
%
%\bibliography{ContactPaper}

\begin{thebibliography}{40}
\bibitem[{foo({\natexlab{a}})}]{foot_Rc}
\bibinfo{note}{For a negligible spin resistance of the ferromagnetic electrodes
  compared to the one of the barrier and to the resistance of the transport
  channel ($R_{sq}$) we call the resistance of the interface barrier between
  the contact electrode and the transport channel ``contact resistance''
  ($R_C$). The contact resistance consists of the parallel resistance for spin
  up ($R_\uparrow$) and spin down ($R_\downarrow$). We assume $R_\uparrow
  \approx R_\downarrow$ as we are discussing polarizations of $P < 30\%$
  leading to an error in determining the effect of the contact resistance on
  the spin relaxation of less than $10 \%$. In the case of transparent contacts
  (i.e. no high resistive barrier), $R_C$ is equal to the spin resistance of
  the ferromagnetic leads $R_{spin}^F$. In this case we also have to replace
  the spin polarization of the barrier $P$ by the spin polarization of the
  ferromagnetic leads $P^F$.}
\bibitem[{foo({\natexlab{b}})}]{foot_2D}
\bibinfo{note}{We are discussing here a 2-dimensional channel like graphene. In
  case of a 3-dimensional channel $R$ would be $R= (R_C/\rho) A$ where $A$ is
  the cross section and $\rho$ the resistivity of the transport channel.}
\bibitem[{foo({\natexlab{c}})}]{foot_Fit}
\bibinfo{note}{To fit the data we first normalize the simulated $R_{nl}$ values
  with $R_{nl}(B=0)$ and keep the fitting amplitude fixed as $1$. This enables
  a more standardized fitting procedure and reduces the fitting time
  considerably. The fixed amplitude can be justified as fitting with a
  non-fixed amplitude gives in most cases a variation of the amplitude by less
  than $1\%$ and only varies more in the regime where $R/\lambda \leq 0.1$ and
  $L/\lambda \leq 0.1$.}
\end{thebibliography}
%

\end{document}